\documentclass[12pt,preprint]{aastex}
\usepackage{rotating}

\def\dfplot#1{\plotone{#1}}
\def\BE{\begin{equation}}
\def\BEL#1{\begin{equation}\label{#1}}
\def\EE{\end{equation}}
\newcommand{\HII}{H\,{\scriptsize II}}

\newcommand{\Halpha}{H$\alpha$}
\newcommand{\etal}{{et al.}~}

\newcommand{\degree}{^\circ}

\begin{document}

\title{Sloan Digital Sky Survey Imaging of Low Galactic Latitude Fields:
Technical Summary and Data Release}

\author{
Douglas P. Finkbeiner\altaffilmark{\ref{Hubble},\ref{Princeton}},
Nikhil Padmanabhan\altaffilmark{\ref{Princetonphys}},
David J. Schlegel\altaffilmark{\ref{Princeton}},
Michael A. Carr\altaffilmark{\ref{Princeton}},
James E. Gunn\altaffilmark{\ref{Princeton}},
Constance M. Rockosi\altaffilmark{\ref{Hubble},\ref{Washington}},
Maki Sekiguchi\altaffilmark{\ref{Kashiwa}},
Robert H. Lupton\altaffilmark{\ref{Princeton}},
G. R. Knapp\altaffilmark{\ref{Princeton}},
\v{Z}eljko Ivezi\'{c}\altaffilmark{\ref{Princeton}},
Michael R. Blanton\altaffilmark{\ref{NYU}},
David W. Hogg\altaffilmark{\ref{NYU}},
Jennifer K. Adelman-McCarthy\altaffilmark{\ref{Fermilab}},
James Annis\altaffilmark{\ref{Fermilab}},
Jeffrey Hayes\altaffilmark{\ref{Catholic}},
Ellynne Kinney\altaffilmark{\ref{APO}},
Daniel C. Long\altaffilmark{\ref{APO}},
Uro\v{s} Seljak\altaffilmark{\ref{Princetonphys}},
Michael A. Strauss\altaffilmark{\ref{Princeton}},
Brian Yanny\altaffilmark{\ref{Fermilab}},
Marcel A. Ag\"ueros\altaffilmark{\ref{Washington}},
Sahar S. Allam\altaffilmark{\ref{NMSU}},
Scott F. Anderson\altaffilmark{\ref{Washington}},
Neta A. Bahcall\altaffilmark{\ref{Princeton}},
Ivan K. Baldry\altaffilmark{\ref{JHU}},
Mariangela Bernardi\altaffilmark{\ref{CMU}},
William N. Boroski\altaffilmark{\ref{Fermilab}},
John W. Briggs\altaffilmark{\ref{Yerkes}},
J. Brinkmann\altaffilmark{\ref{APO}},
Robert J. Brunner\altaffilmark{\ref{Illinois}},
Tam\'as Budav\'ari\altaffilmark{\ref{JHU}},
Francisco J. Castander\altaffilmark{\ref{Barcelona}},
Kevin R. Covey\altaffilmark{\ref{Washington}},
Istv\'an Csabai\altaffilmark{\ref{Eotvos},\ref{JHU}},
Mamoru Doi\altaffilmark{\ref{IoAUT}},
Feng Dong\altaffilmark{\ref{Princeton}},
Daniel J. Eisenstein\altaffilmark{\ref{Arizona}},
Xiaohui Fan\altaffilmark{\ref{Arizona}},
Scott D. Friedman\altaffilmark{\ref{STScI}},
Masataka Fukugita\altaffilmark{\ref{ICRRUT}},
Bruce Gillespie\altaffilmark{\ref{APO}},
Eva K. Grebel\altaffilmark{\ref{MPIA}},
Vijay K. Gurbani\altaffilmark{\ref{Fermilab},\ref{Lucent2}},
Ernst de Haas\altaffilmark{\ref{Princeton}},
Frederick H. Harris\altaffilmark{\ref{NOFS}},
John S. Hendry\altaffilmark{\ref{Fermilab}},
Gregory S. Hennessy\altaffilmark{\ref{USNO}},
Sebastian Jester\altaffilmark{\ref{Fermilab}},
David E. Johnston\altaffilmark{\ref{Princeton}},
Anders M. Jorgensen\altaffilmark{\ref{LANL}},
Mario Juri\'{c}\altaffilmark{\ref{Princeton}},
Stephen M. Kent\altaffilmark{\ref{Fermilab}},
Alexei Yu. Kniazev\altaffilmark{\ref{MPIA}},
Jurek Krzesinski\altaffilmark{\ref{APO},\ref{MSO}},
R. French Leger\altaffilmark{\ref{Fermilab}},
Huan Lin\altaffilmark{\ref{Fermilab}},
Jon Loveday\altaffilmark{\ref{Sussex}},
Ed Mannery\altaffilmark{\ref{Washington}},
David Mart\'{\i}nez-Delgado\altaffilmark{\ref{MPIA}},
Peregrine M. McGehee\altaffilmark{\ref{NMSU},\ref{LANL2}},
Avery Meiksin\altaffilmark{\ref{Edinburgh}},
Jeffrey A. Munn\altaffilmark{\ref{NOFS}},
Eric H. Neilsen, Jr.\altaffilmark{\ref{Fermilab}},
Peter R. Newman\altaffilmark{\ref{APO}},
Atsuko Nitta\altaffilmark{\ref{APO}},
George Pauls\altaffilmark{\ref{Princeton}},
Thomas R. Quinn\altaffilmark{\ref{Washington}},
R. R. Rafikov\altaffilmark{\ref{IAS}},
Gordon T. Richards\altaffilmark{\ref{Princeton}},
Michael W. Richmond\altaffilmark{\ref{RIT}},
Donald P. Schneider\altaffilmark{\ref{PSU}},
Joshua Schroeder\altaffilmark{\ref{Princeton}},
Kazu Shimasaku\altaffilmark{\ref{DoAUT}},
Walter A. Siegmund\altaffilmark{\ref{Hawaii}},
J. Allyn Smith\altaffilmark{\ref{Wyoming},\ref{LANL}},
Stephanie A. Snedden\altaffilmark{\ref{APO}},
Albert Stebbins\altaffilmark{\ref{Fermilab}},
Alexander S. Szalay\altaffilmark{\ref{JHU}},
Gyula P. Szokoly\altaffilmark{\ref{MPIEP}},
Max Tegmark\altaffilmark{\ref{Penn}},
Douglas L. Tucker\altaffilmark{\ref{Fermilab}},
Alan Uomoto\altaffilmark{\ref{JHU},\ref{CarnegieObs}},
Daniel E. Vanden Berk\altaffilmark{\ref{Pitt}},
David H. Weinberg\altaffilmark{\ref{OSU}},
Andrew A. West\altaffilmark{\ref{Washington}},
Naoki Yasuda\altaffilmark{\ref{ICRRUT}},
D. R. Yocum\altaffilmark{\ref{Fermilab}},
Donald G. York\altaffilmark{\ref{Chicago},\ref{EFI}},
Idit Zehavi\altaffilmark{\ref{Arizona}}
}
\altaffiltext{1}{
Hubble Fellow
\label{Hubble}}
\altaffiltext{2}{
Department of Astrophysical Sciences, Princeton University, Princeton, NJ
08544
\label{Princeton}}
\altaffiltext{3}{
Joseph Henry Laboratories, Princeton University, Princeton, NJ
08544
\label{Princetonphys}}
\altaffiltext{4}{
Department of Astronomy, University of Washington, Box 351580, Seattle, WA
98195
\label{Washington}}
\altaffiltext{5}{
Japan Participation Group, c/o 
Institute for Cosmic Ray Research, University of Tokyo, 5-1-5 Kashiwa, Kashiwa City, Chiba 277-8582, Japan
\label{Kashiwa}}
\altaffiltext{6}{
Center for Cosmology and Particle Physics,
Department of Physics,
New York University,
4 Washington Place,
New York, NY 10003
\label{NYU}}
\altaffiltext{7}{
Fermi National Accelerator Laboratory, P.O. Box 500, Batavia, IL 60510
\label{Fermilab}}
\altaffiltext{8}{
Department of Physics, Catholic University of America, Washington DC 20064
\label{Catholic}}
\altaffiltext{9}{
Apache Point Observatory, P.O. Box 59, Sunspot, NM 88349
\label{APO}}
\altaffiltext{10}{
New Mexico State University, Department of Astronomy, P.O. Box 30001, Dept
4500, Las Cruces, NM 88003
\label{NMSU}}
\altaffiltext{11}{
Center for Astrophysical Sciences, Department of Physics \& Astronomy, Johns
Hopkins University, Baltimore, MD 21218
\label{JHU}}
\altaffiltext{12}{
Department of Physics, Carnegie Mellon University, Pittsburgh, PA 15213
\label{CMU}}
\altaffiltext{13}{
Yerkes Observatory, University of Chicago, 373 W. Geneva St., Williams
Bay, WI 53191
\label{Yerkes}}
\altaffiltext{14}{
Department of Astronomy,
University of Illinois,
1002 W. Green Street, Urbana, IL 61801
\label{Illinois}}
\altaffiltext{15}{
Institut d'Estudis Espacials de Catalunya/CSIC, Gran Capita 2-4,
08034 Barcelona, Spain
\label{Barcelona}}
\altaffiltext{16}{
Department of Physics, E\"{o}tv\"{o}s University, Budapest, Pf.\ 32,
Hungary, H-1518
\label{Eotvos}}
\altaffiltext{17}{
Institute of Astronomy, School
of Science, University of Tokyo,
 2-21-1 Osawa, Mitaka, Tokyo 181-0015, Japan
\label{IoAUT}}
\altaffiltext{18}{
Steward Observatory, 933 N. Cherry Ave, Tucson, AZ 85721
\label{Arizona}}
\altaffiltext{19}{
Space Telescope Science Institute, 3700 San Martin Drive, Baltimore, MD
21218
\label{STScI}}
\altaffiltext{20}{
Institute for Cosmic Ray Research, University of Tokyo, 5-1-5 Kashiwa,
Kashiwa City, Chiba 277-8582, Japan
\label{ICRRUT}}
\altaffiltext{21}{
Max-Planck Institute for Astronomy, K\"onigstuhl 17, D-69117 Heidelberg,
Germany
\label{MPIA}}
\altaffiltext{22}{
Lucent Technologies, 2000 Lucent Lane, Naperville, IL 60566
\label{Lucent2}}
\altaffiltext{23}{
US Naval Observatory, 3540 Mass Ave NW, Washington, DC 20392
\label{USNO}}
\altaffiltext{24}{
ISR-4, MS D448, Los Alamos National Laboratory, Los Alamos, NM 87545
\label{LANL}}
\altaffiltext{25}{
Mt. Suhora Observatory, Cracow Pedagogical University, ul. Podchorazych 2,
30-084 Cracow, Poland
\label{MSO}}
\altaffiltext{26}{
Astronomy Centre, University of Sussex, Falmer, Brighton BN1 9QJ, United
Kingdom
\label{Sussex}}
\altaffiltext{27}{
SNS-4, MS H820, Los Alamos National Laboratory, Los Alamos, NM 87545
\label{LANL2}}
\altaffiltext{28}{
Institute for Astronomy,
Royal Observatory,
Blackford Hill,
Edinburgh EH9 3HJ,
Scotland
\label{Edinburgh}}
\altaffiltext{29}{
U.S. Naval Observatory, Flagstaff Station, P.O. Box 1149, Flagstaff, AZ
86002-1149
\label{NOFS}}
\altaffiltext{30}{
Institute for Advanced Study, Olden Lane, Princeton, NJ 08540
\label{IAS}}
\altaffiltext{31}{
Physics Department, Rochester Institute of Technology, 84 Lomb Memorial
Drive, Rochester, NY 14623-5603
\label{RIT}}
\altaffiltext{32}{
Institute for Astronomy, 2680 Woodlawn Road, Honolulu, HI 96822
\label{Hawaii}}
\altaffiltext{33}{
Department of Astronomy and Astrophysics, the Pennsylvania State
University, University Park, PA 16802
\label{PSU}}
\altaffiltext{34}{
Department of Astronomy and Research Center for the Early Universe, School
of Science, University of Tokyo,
 7-3-1 Hongo, Bunkyo, Tokyo 113-0033, Japan
\label{DoAUT}}
\altaffiltext{35}{
University of Wyoming, Dept. of Physics \& Astronomy, Laramie, WY 82071
\label{Wyoming}}
\altaffiltext{36}{
Max-Planck-Institut f\"ur extraterrestrische Physik, 
Giessenbachstrasse 1, D-85741 Garching, Germany
\label{MPIEP}}
\altaffiltext{37}{
Department of Physics, University of Pennsylvania, Philadelphia, PA 19104
\label{Penn}}
\altaffiltext{38}{
Carnegie Observatories,
813 Santa Barbara St.,
Pasadena, CA  91101
\label{CarnegieObs}}
\altaffiltext{39}{
Department of Physics and Astronomy, University of Pittsburgh, 3941 O'Hara
St., Pittsburgh, PA 15260
\label{Pitt}}
\altaffiltext{40}{
Department of Astronomy, Ohio State University, Columbus, OH 43210
\label{OSU}}
\altaffiltext{41}{
Department of Astronomy and Astrophysics, The University of Chicago, 5640 S.
Ellis Ave., Chicago, IL 60637
\label{Chicago}}
\altaffiltext{42}{
Enrico Fermi Institute, The University of Chicago, 5640 S. Ellis Ave.,
Chicago, IL 60637
\label{EFI}}

\begin{abstract}
The Sloan Digital Sky Survey (SDSS) mosaic camera and telescope have
obtained five-band optical-wavelength imaging near the Galactic plane
outside of the nominal survey boundaries.
These additional data were obtained during commissioning and
subsequent testing of the SDSS observing system, and they provide
unique wide-area imaging data in regions of high obscuration and star
formation, including numerous young stellar objects, Herbig-Haro
objects and young star clusters.
Because these data are outside the Survey regions in the Galactic
caps, they are not part of the standard SDSS data releases.
This paper presents imaging data for 832 square degrees of sky
(including repeats), in the star-forming regions of Orion, Taurus, and
Cygnus.  About 470 square degrees are now released to the public, with
the remainder to follow at the time of SDSS Data Release 4. 
The public data in Orion include the star-forming region NGC 2068/NGC
2071/HH24 and a large part of Barnard's loop. 

\emph{Subject headings: }
atlases ---
catalogs ---
surveys
\end{abstract}

\section{INTRODUCTION}
\label{sec_intro}
The Sloan Digital Sky Survey (SDSS) is a 5-band photometric survey of
8500 square degrees of the Northern sky and a concurrent
redshift survey of nearly a million galaxies and 100,000 quasars
selected from the imaging survey \cite{york00}.  The primary
purpose of the project is to investigate the
large scale structure of the Universe and pursue other extragalactic science.
The official survey region therefore largely lies above Galactic latitude 
$|b| > 30\degree$ and was carefully chosen to minimize the effect of the
troublesome dust near the Galactic plane.
This naturally excludes some of the most beautiful parts of the sky,
and those areas of most interest for Galactic science.

However, a significant amount of imaging data \emph{has}
been obtained at low Galactic latitude by SDSS, both during
telescope commissioning and for calibration at sidereal times
when the main survey region was unavailable.  During commissioning,
the SDSS camera \cite{gunn98} was
operational before the telescope control system was stable.  The
camera data acquisition is carried out in drift-scan or ``Time Delay
\& Integrate'' (TDI)
mode, with the camera crossing the sky at the sidereal rate; the large
field of view ($3^{\circ}$) mandates that this be done along great
circles.  Accordingly, much of the commissioning work was done by
parking the telescope at the Celestial Equator and drift-scanning the
sky. During commissioning observations in Fall 1998 (when
the South Galactic Polar Cap is available for observation), the drift
scanning continued outside the SDSS survey area and passed through
NGC 2068 and NGC 2071 in Orion.  Many
of the commissioning runs produced data of high
scientific quality, some of which are part of
the SDSS Early Data Release \cite[][hereafter EDR]{stoughton02}, where a
detailed description of the imaging data can be found.  
SDSS Runs 259 and 273 generated targets for spectroscopic commissioning,
resulting in the first extremely metal-poor galaxy found by SDSS
\cite{kniazev03}, but have never been made public. 
Since 1999,
additional imaging data at low Galactic latitudes have been obtained
for a variety of testing and calibration purposes.  However, none of these
low-latitude data,
external to the SDSS survey area, are included in the
SDSS data releases (to date the EDR; the First Data Release, DR1,
Abazajian \etal\ 2003; and DR2, Abazajian \etal\ 2004).  
We have reduced and calibrated these data as part of a re-reduction
and re-calibration of all the imaging data
(``ubercalibration'', Schlegel et al. 2004), and we describe these
data in the present paper.

The areas of sky observed are described in Section 2.  The data
products described herein are very similar to those
produced by the SDSS, but differ in photometric calibration and data
format.  These differences are described in Section 3, and some
examples of science applications are briefly discussed in Section 4.
The data from 1998-1999 are publicly released
with this paper and may be accessed via the WWW.\footnote{
\texttt{http://photo.astro.princeton.edu}}
Future data releases will also be accessible at this site. 

\section{THE DATA}
\label{sec_data}

\subsection{Sky Coverage}
\label{sec_coverage}

\begin{figure}[tb]
\epsscale{1.0}
\dfplot{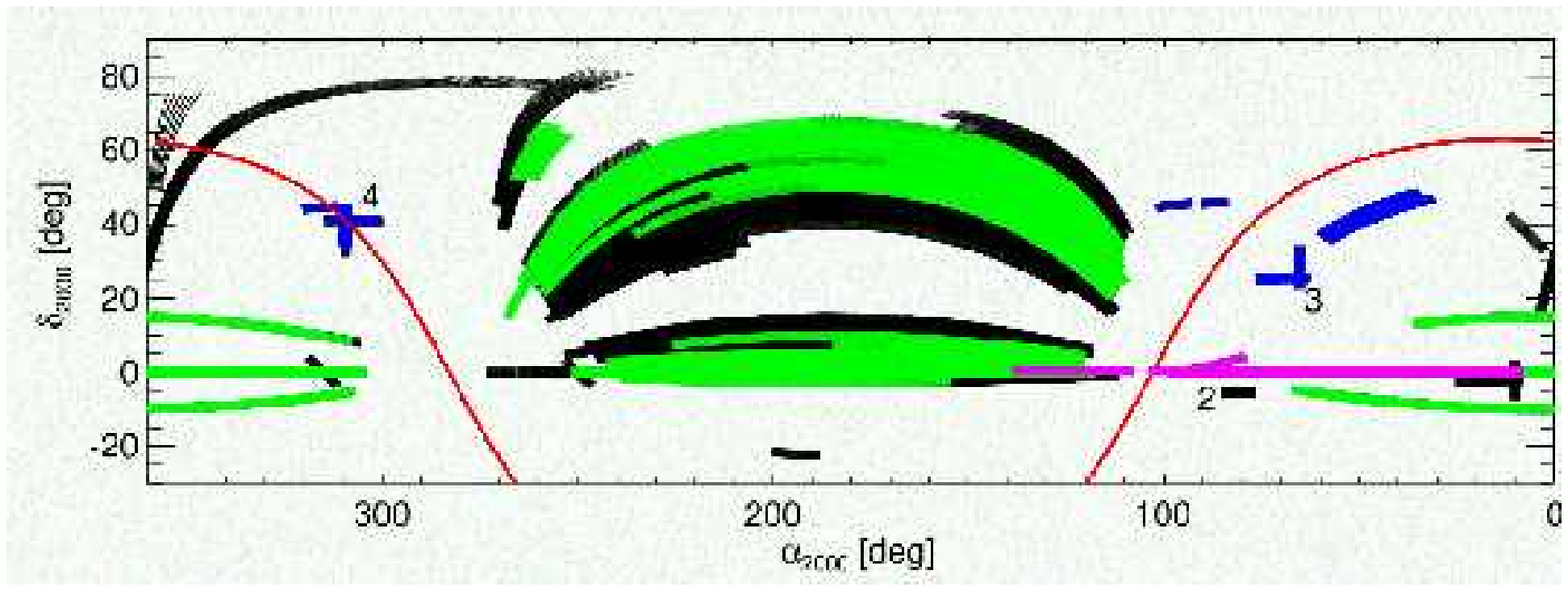}
\figcaption{
The SDSS footprint in equatorial coordinates, consisting of the
(currently released) Orion
data (\emph{magenta}), other Galactic plane runs (\emph{blue}), SDSS
DR2 (\emph{green}), and other SDSS data taken through December, 2003
(\emph{black}).  Numbers mark the regions shown in Figures
\ref{fig_orion}-\ref{fig_cygnus}.
\label{fig_overview}
}
\end{figure}

The sky coverage of the imaging runs made to date outside the SDSS area is
given in Table \ref{tab_runlist} and Figure \ref{fig_overview}.
The figure shows
the footprint of the survey (including these runs) in equatorial
coordinates and indicates the approximate locations of images in
Figures \ref{fig_orion},\ref{fig_taurus}, and \ref{fig_cygnus}.
 Table \ref{tab_runlist} provides information about the
location and image quality of each run.  An imaging run consists of
six long images in each filter, each the width
of one CCD ($13.52'$ on the sky) 
and separated by slightly less than one CCD width ($11.65'$), produced by 
the six columns of CCDs in the mosaic camera.  A run may last the
entire night and be over $100\degree$ long, or can be shorter. 
A \emph{strip} is the area covered by the six camera columns from one
survey pole to the other;
 a \emph{stripe} is a pair of interleaving \emph{strips} and
 completely covers a $2.5\times180 \degree$ stripe on the sky. 
The imaging data are divided 
into $8.98'\times13.52'$ {\it frames} for further processing; the aligned 
frames in the five bands are called a {\it field}.  Since the sky
passes through the u,g,r,i,z SDSS filters in succession (in order r,i,u,z,g)
there is a ramp-up time corresponding to 10 frames at the start of
each imaging run. 
Focus and tracking adjustments are made at the beginning of each
run, adding variable amounts of overhead to the ramp-up time.
The range of fields for each run given in Table \ref{tab_runlist}
define the range of useful data.  A detailed description of the
observing procedures is given by Gunn \etal (1998) and York \etal\ (2000).

The imaging runs comprising the ``Orion'' data set, released in the 
present paper, were made in fall 1998 and fall 1999.  As Table
\ref{tab_runlist}
and Figure \ref{fig_overview} show, there have been quite a few 
low-latitude observations since then, usually for system checking and
calibration.  Non-photometric data are included, because proper motion
studies can make use of the astrometry even in unphotometric runs.
The data listed in Table \ref{tab_runlist} are those which, after
reduction and processing, prove to be of science quality, and the
photometric reliability is indicated.

\subsection{Processing}
\label{sec_processing}

The SDSS photometric pipeline consists of four sequential steps:
\texttt{ssc} and \texttt{psp} (point-spread function estimation),
\texttt{astrom} \cite[][astrometry]{pier03}, and \texttt{frames}
\citep[object identification, deblending, and photometry;][and
in preparation]{lupton01,stoughton02,lupton03}.  These pipelines run
with little human intervention.
The photometric
pipeline corrects the imaging data by interpolating over defects such as
bad columns and cosmic rays; provides flat-field, photometric 
and astrometric calibration; and identifies, deblends, measures and classifies 
objects.  The resulting outputs consist of corrected frames, atlas image
cutouts for every object, and a catalog of object positions, magnitudes
and image classifications in the five SDSS bands.  The catalog outputs
also include flags which describe the image processing, including
whether the object contains any saturated pixels, whether bad data were
interpolated over, and whether the object was deblended.  
Careful attention to these processing flags is critical for proper
scientific use of the data.
A complete
description of the data outputs can be obtained at the above website$^1$, 
at the SDSS DR2 web site\footnote{
\texttt{http://www.sdss.org}}, and in
papers \cite{stoughton02,abazajian03,abazajian04}.

The \texttt{frames} pipeline produces a catalog of objects
(\texttt{fpObjc} files) described by instrumental quantities: $(x,y)$
positions on a frame, CCD counts, and radii.  
Astrometric calibrations are applied
by the astrometric pipeline (\texttt{astrom}) and are accurate to
better than $0.1''$ RMS in each coordinate \cite{pier03}.
Because early commissioning data sometimes had poor telescope
pointing information (pointing errors of up to $0.5\degree$),
astrometric pre-processing using the discrete cross-correlation method
described by Hogg et al. (2001)
has been performed to allow such data to be
processed in an automated fashion.
The flat-fielding and photometric calibration were derived from
a global recalibration using all of the available repeat imaging
(Schlegel et al.\ 2004).
The zero-points of the photometric solutions are forced to agree
on average with the calibrations derived from the Photometric Telescope
\cite{fukugita96,york00,smith02} used in SDSS DR2.

\subsubsection{Automation with {\rm\texttt{photoop}}}
The importance of a completely automated data processing pipeline
cannot be overemphasized; it ensures that all the reductions are
carried out in a homogeneous fashion and protects against
human biases. Furthermore, it ensures that all reductions
are perfectly reproducible, allowing comparisons between different
reduction attempts. This is a necessary prerequisite for a survey like
the SDSS that targets subsamples of galaxies for follow up
observations because it enables one to measure the completeness of these
subsamples even as the underlying software evolves.

While the standard SDSS pipelines perform well within the standard
survey region, the radically different characteristics of the data
near the Galactic plane require adjustments of various pipeline
parameters to process these data. In order to implement these in an
automated manner, we have developed a meta-pipeline, \texttt{photoop},
to manage all aspects of processing.  In production mode, the only
human intervention required is to specify the runs to be processed;
after that, \texttt{photoop} sets up the photometric pipeline and
generates the required input data for each run, including flat-fields,
astrometric catalogs, and configuration files.  Perl scripts then
manage the actual processing across several UNIX servers with a shared
NFS file system.  Other \texttt{photoop} scripts automate daily
maintenance tasks: monitoring the processing status of each run,
posting results to an internal web page, and testing data integrity.
Notices of pipeline failures and data integrity warnings are
automatically sent by email to responsible parties. 
Furthermore, \texttt{photoop} has been designed to be
portable so that anyone with sufficient computing power and access to
the raw data can replicate the entire processing system.

\subsubsection{Processing Software Versions and Reruns}
As the SDSS software pipelines have evolved over time, it has been
necessary to keep track of the various processing attempts with
\emph{rerun} numbers.  Data processing done via \texttt{photoop} is
strictly versioned, with rerun numbers $> 100$ to avoid confusion with
the main Survey rerun numbers ($0-99$).  There is a one-to-one
correspondence between \emph{rerun} number and software versions.  The
current rerun as of this writing is 137, corresponding to \texttt{ssc}
v5\_3\_4, \texttt{astrom} v3\_7, \texttt{photo} v5\_4\_25, and v4
flatfields and \texttt{photoop} v1\_0.  For historical reasons,
the \texttt{psp} and \texttt{frames} steps of the pipeline are both
part of the \texttt{photo} software product.
The instructions below for accessing and using the data
refer to rerun 137.  Future refinements of the software will
have higher rerun numbers and will be announced on the website$^1$.

\subsubsection{Calibration}
\label{sec_calib}

Survey data releases are indirectly calibrated to a set of primary
photometric standard stars measured the US Naval Observatory (USNO) 1m 
telescope
in the SDSS filters.  This photometric system is denoted by
u',g',r',i',z' and is defined by Smith \etal\ (2002).  These primary
standards are bright enough to saturate the 2.5m survey telescope, so
the photometric system is transfered to a set of secondary standard
star ``patches'' observed by a 20-inch ``Photometric Telescope'' (PT)
situated adjacent to the survey telescope.  These secondary standards
are used to calibrate the 2.5m native filter system (denoted
u,g,r,i,z) to the USNO system for typical star colors.  Note that the
u',g',r',i',z' and u,g,r,i,z systems are significantly different (see
e.g. Stoughton et al. 2002).
PT patches are sparse or
nonexistent for much of the Orion region, so we have instead used the
ubercalibration algorithm (Schlegel \etal\ 2004) to tie the eight Orion
runs together photometrically.  This algorithm lets the calibration
zero-points in each CCD ($a$-terms) and atmospheric
extinction ($k$-terms) float night by night, constrained by multiple
observations of stars in run overlaps. 
Run 308 does not overlap any of the other Orion runs, and is therefore
calibrated using the same $a$- and $k$-terms as run 307 from the same
night.
Ubercalibration minimizes
the RMS magnitude residuals in repeat observations of $\sim10^6$ stars
by adjusting these $a$- and $k$-terms.  Because the north and south
strips of the equatorial stripe overlap only at the camera column
edges, and the flatfields are
less certain there, it is necessary to use a
perpendicular scan (run 2766) to tie the 12 independent camera
columns together.  This run is connected to the Orion runs via four other
equatorial runs (94, 125, 1755, 2677).
For equatorial drift scans the airmass is constant, making the $a$-
and $k$-terms degenerate, so $k$-terms are fixed to canonical values. 

\begin{figure}[tb]
\epsscale{1.0}
\dfplot{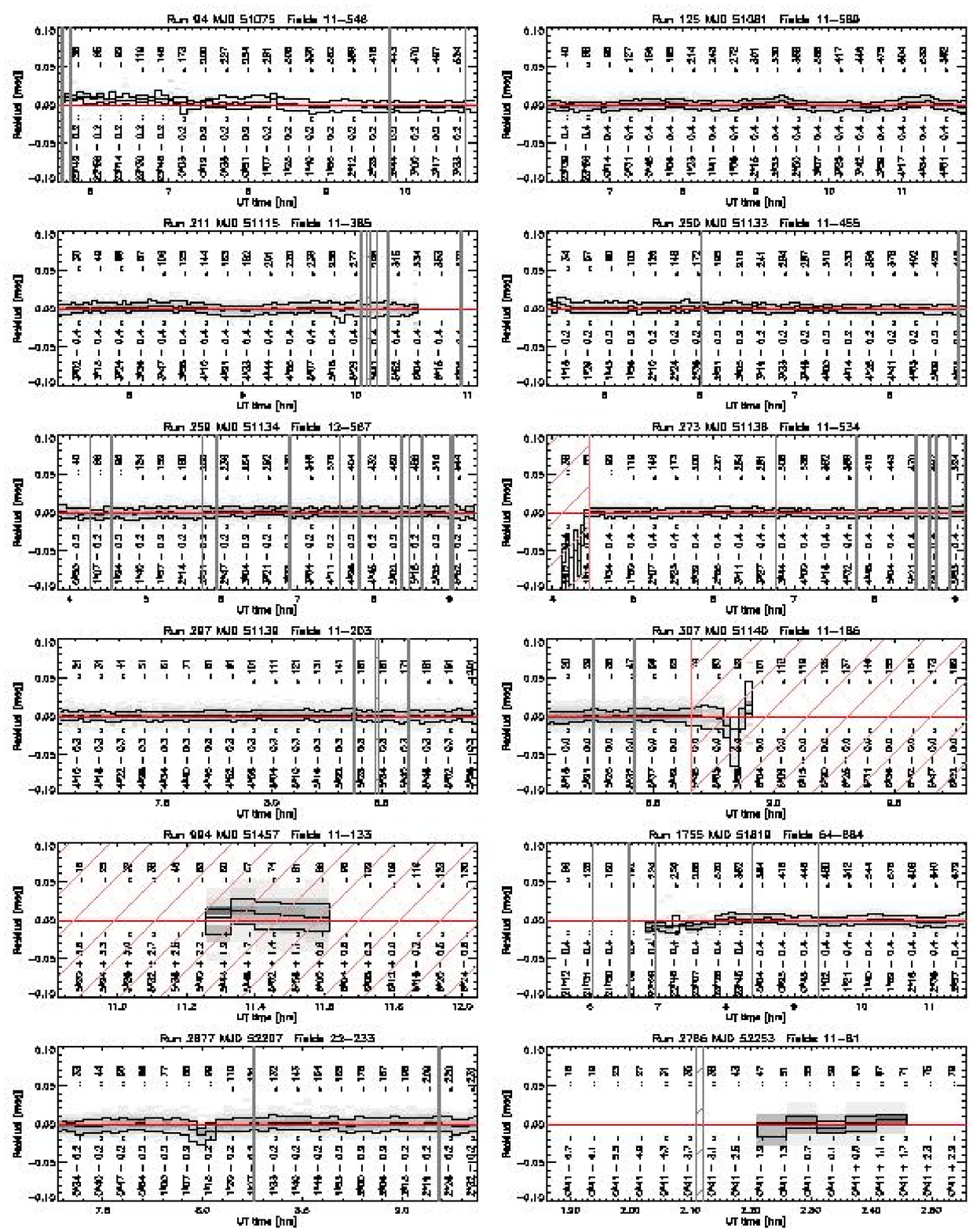}
\figcaption{
Measured magnitudes minus means for each star of each run used in the
Orion ubercalibration (see \S\ref{sec_calib}).  Stars brighter than
(19.0,19.0,19.0,18.5,17.0) in (u,g,r,i,z) are used in the calibration;
results are shown for $r$ band only.  Differences are
represented as gray scales with 25th, 50th, and 75th percentile
lines.  Results are similar for g and i bands, with more photometric
scatter in u and z bands (see Fig. \ref{fig_ubercal_hist}).  Cloudy or
otherwise unphotometric data are indicated by red cross-hatches,
and gray hatches indicate fields where the pipeline was unable to
determine a good point spread function;
these data are
not used in fits.  Run 308 does not appear here because it has no
overlap with the others, but is calibrated by fixing its parameters to
run 307. 
\label{fig_ubercal_resid}
}
\end{figure}

\begin{figure}[tb]
\epsscale{1.0}
\dfplot{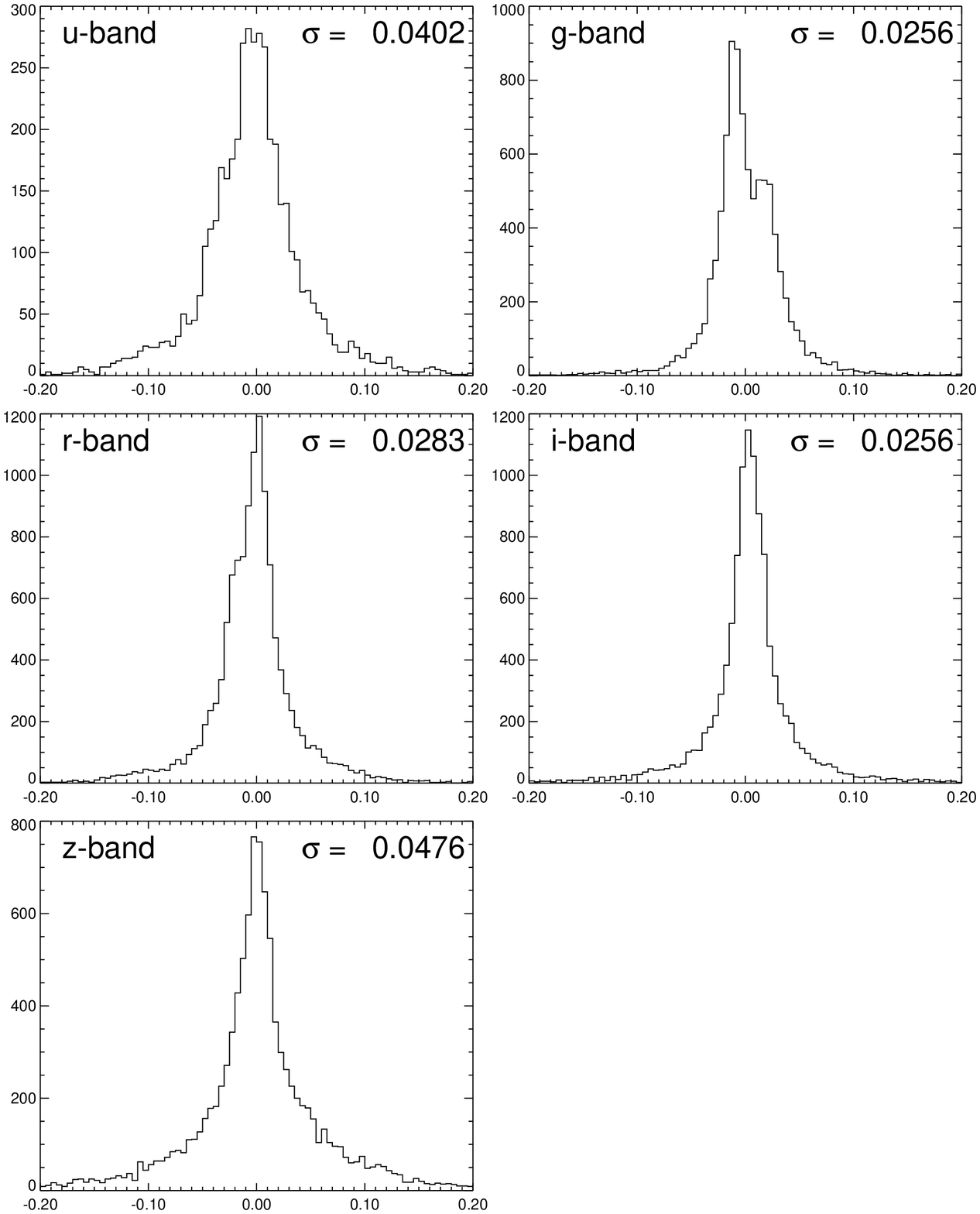}
\figcaption{
Histograms of aperture magnitude differences between runs 259 and 297
for the 5 SDSS bands.  Stars with $7.43''$ radius aperture magnitudes
brighter than (19.0,19.0,19.0,18.5,17.0)
in (u,g,r,i,z) are shown.  Note that the measurement uncertainty in
each run is a factor of $\sqrt2$ smaller than the histogram width.
The bimodality in the g-band is likely
caused by $1-2\%$ variation in the flat-fields with time.
\label{fig_ubercal_hist}
}
\end{figure}

Most stars are observed multiple times by this set of
runs, and for each star the residual between each run and the mean of
all observations of that star is shown in Figure
\ref{fig_ubercal_resid}.  Only stars with $7.43''$ radius aperture
magnitudes brighter than (19.0,19.0,19.0,18.5,17.0) in (u,g,r,i,z)
were used for the calibration, and the resulting difference
histograms have a $5\sigma$ clipped RMS of (40,26,28,26,48)
millimag (Figure \ref{fig_ubercal_hist}).  This calibration method
will be described in detail elsewhere (Schlegel \etal\ 2004) and
astrophysical tests of its accuracy will be provided by Finkbeiner
\etal\ (2004).

\subsection{Public Data Access}

Table \ref{tab:products} lists all software and ancillary data
products \cite[in addition to the standard survey
software described by][]{stoughton02} that are used in the data
processing, access and analysis steps. 
These products are organized with the Concurrent Versions System
(CVS\footnote{\texttt{http://www.cvshome.org}}).
IDL\footnote{\texttt{http://www.rsinc.com}} routines for displaying
and calibrating images, extracting photometric parameters for sources,
and matching sources to other public catalogs are included.  Recipes
using some of the most powerful routines are given in examples below.
We refer the reader to the website$^1$ for details
on downloading and installing these packages; here we limit ourselves to brief
descriptions of the data products that make up this data release.

This data release contains three datasets: calibrated images for every
field, object catalogs for every field, and trimmed catalogs of ``stars''
(all point sources, including quasars) and ``galaxies'' (extended sources)
for every run. The images have had cosmic rays
removed, CCD defects corrected, and have been flatfielded and
photometrically calibrated.  In addition, these images have
accurate astrometry
stored in a FITS-compliant header and therefore do not depend on any
auxiliary information.  The object catalogs (\texttt{calibObj}) include
all data fields in the uncalibrated
\texttt{fpObjc} catalogs produced by the SDSS \texttt{frames} pipeline,
as well as calibrated quantities.  Object fluxes are
calibrated and the CCD positions of
objects are translated to equatorial coordinates using the best
fit astrometric solution from the SDSS \texttt{astrom} pipeline. An overview
of the differences between these \texttt{calibObj} catalogs and the
standard SDSS \texttt{tsObj} files 
(available at the Data Archive
Server\footnote{\texttt{http://das.sdss.org/DR2/data/imaging/}; see
  Stoughton et al. 2002 for more details})
is given in \S \ref{sec_calibobj}, and
the \texttt{calibObj} format is described in detail in Appendix
\ref{app_calibobj}.

The \texttt{calibObj} catalogs for the 8 public runs require 50 GB.
For users who prefer data in a more compact format, we also
provide trimmed stellar and galaxy catalogs for each camera column in
each of the 8 runs, which total 1.5 GB for the stars and 2.5 GB for the
galaxies. The trimmed catalogs contain all stars with
any of $(u,g,r,i,z)$ brighter than $(22.5, 22.5, 22.5, 22.0, 21.5)$ mag
respectively, and galaxies brighter than $(21.0, 22.0, 22.0, 20.5,$ $
20.1)$.  These cuts were made after applying the Schlegel, Finkbeiner,
\& Davis (1998, hereafter SFD98) extinction correction for the
purpose of object selection -- the extinction correction is
\emph{not} applied to the flux values in the catalogs.  These trimmed
star and galaxy catalogs are a
strict subset of the quantities described in Appendix
\ref{app_calibobj}, and contain no additional
information.  The list of fields contained is given on the website.$^1$

Instructions for downloading any of these data are at the website. We
advise users to organize downloaded data using the directory structure
described there,
as this ensures compatibility with our released software.

\subsection{Release Schedule}
Runs 211 through 308, comprising 470 square degrees, are now
available to the public at the website$^1$.
Further data will
be released on a schedule which approximately parallels that of the
SDSS data releases.  We plan to release runs through 4119 at
the time of the SDSS Data Release 4, expected in 2005.
Note that these data, unlike the SDSS data releases, contain no
spectra.


\section{Data Formats}
\label{sec_calibobj}

The Galactic plane data processing and data formats differ somewhat from those
of SDSS DR2, partly from changes necessary to process these data, and
partly to rationalize some naming conventions. The catalogs, produced 
by the current data processing, are in
\texttt{calibObj} files, whose format we now describe.

\begin{itemize}

\item All data fields output by the \texttt{frames} step of the
photometric pipeline in \texttt{fpObjc} files are retained, and not
overwritten as they are in \texttt{tsObj} files.

\item Object fluxes are given in linear units, instead of the asinh
  magnitudes \cite[or ``Luptitudes,''][]{lupgunn} used in the other
  data releases.  This facilitates e.g. coadding of repeat imaging at the
  catalog level. 

\item Calibrated quantities are presented in nano-Maggies (abbreviated
nMgy, a flux density), where 1 Mgy is the AB flux density of a 0th
magnitude flat-spectrum object (Oke \& Gunn 1983).  An AB magnitude of
22.5 corresponds to
1 nMgy in any filter.  The approximate conversion to physical units is
1 Mgy = 3631
Jy where 1 Jy = $10^{-26}{\rm~W~m^{-2}~Hz^{-1}}$ = $10^{-23} {\rm
erg~s^{-1}~cm^{-2}~Hz^{-1}}$.  This conversion is only accurate to
about 5\% \cite{oke83}, although relative zero-point offsets among the
SDSS filters are determined to $1-2\%$ (Abazajian et al
2004).  Note that the
conversion of a measured calibrated
magnitude of a source in a broad-band filter system to a physical flux
density depends on the spectral shape of the source, so
precise conversion factors do not exist.

%
\item Names are rationalized.  In \texttt{fpObjc} files, for historical reasons,
model counts are known as COUNTS\_MODEL and Petrosian counts as
PETROCOUNTS.  In the standard survey \texttt{tsObj}
files these fields are overwritten with
calibrated magnitudes, causing confusion.  In the \texttt{calibObj}
data structure, these uncalibrated data fields appear unchanged, with
the corresponding calibrated quantities appended as MODELFLUX and
PETROFLUX.

\item Flux uncertainties are nearly Gaussian in the low signal to
noise limit (unlike errors in magnitudes).  We express these errors as
an inverse variance (ivar).  This is convenient for inverse variance
weighting when combining quantities, and handles zero signal/noise
(ivar = 0) gracefully, as demonstrated in Appendix
\ref{app_ivar}. Names are e.g. PETROFLUX\_IVAR.

\item Extinction in magnitudes (from SFD98) is given in the five SDSS
bands, and is
  called EXTINCTION, not REDDENING
  as in the \texttt{tsObj} files.  
  These values are calculated assuming the standard $R_V=3.1$
  reddening law \cite{ccm89}, which may be inappropriate for some
  low-latitude regions.

\item Model profile angles such as PHI\_ISO\_DEG, PHI\_DEV\_DEG,
PHI\_EXP\_DEG are expressed as angles in degrees East of
North.

\item PSF\_FWHM (arcsec) is given for every object in each band.  Due to the
  telescope optics and seeing, the PSF varies across a frame. The
  PSF is modeled by a set of eigenfunctions measured by bright stars
  in the frame, and the interpolated PSF calculated at the position of 
  every object detected by \texttt{photo}.  Note that this same
  information is available via the adaptive moments of the
  reconstructed PSF.

\item Wherever possible, objects are matched to the FIRST Radio Catalog 
\cite{becker95}, the 2MASS Point Source Catalog \cite{cutri03} and 
the USNO-B astrometric survey \cite{monet03}.

\end{itemize}


\section{Science}
\label{sec_science}

The science enabled by the wide area imaging reported here generally
lies in the areas of Galactic structure, star formation and interstellar
matter.  Several examples are noted here; for the simplest of these,
we also provide fragments of IDL code that illustrate the
functionality of our publicly available software (Table \ref{tab:products}).

\subsection{Displaying an image}
\label{sec_atv}

\begin{figure}[tb]
\epsscale{0.8}
\dfplot{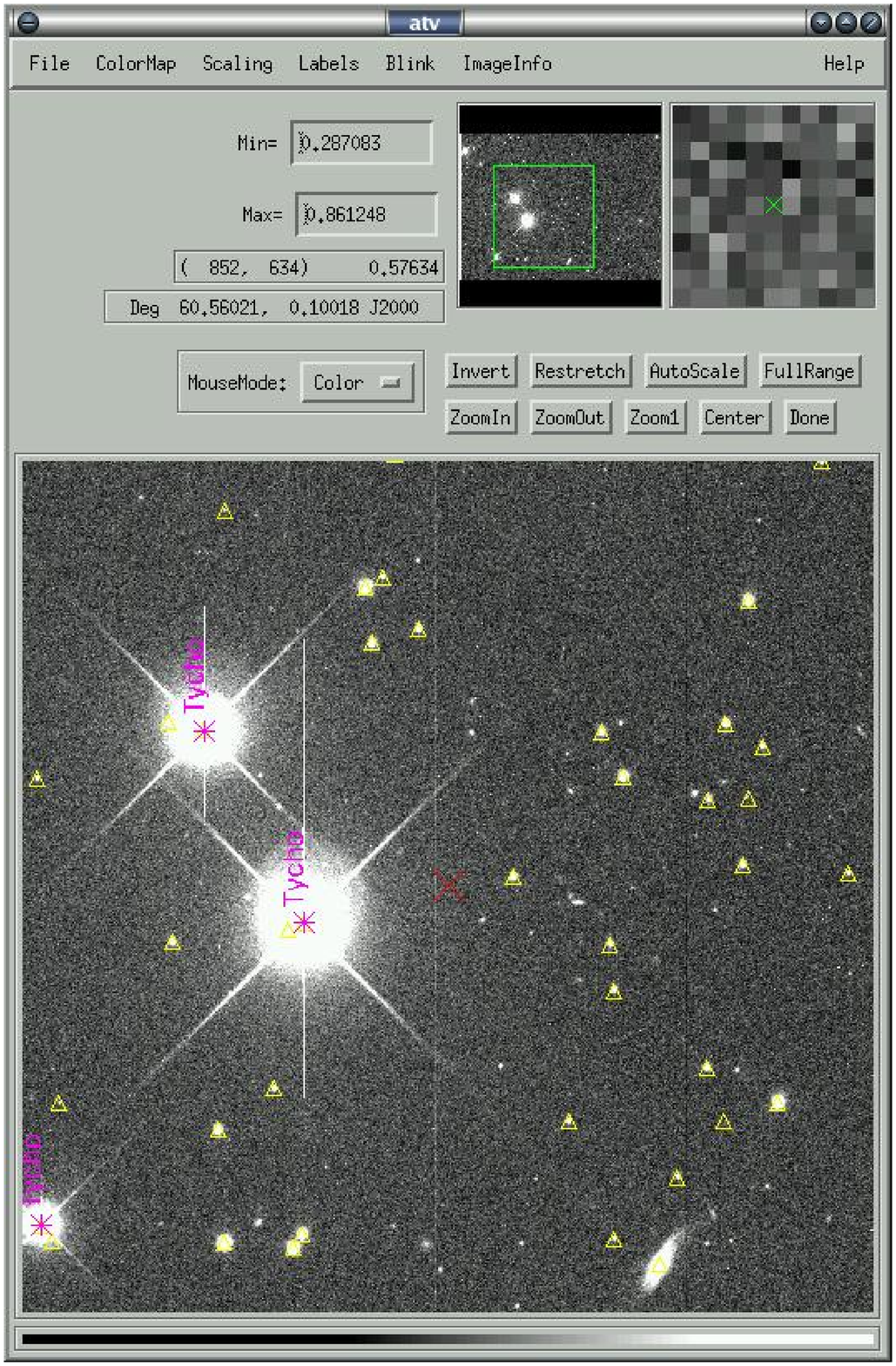}
\figcaption{
The display tool \texttt{atvsdss} finds a raw SDSS
    image at the
given coordinates, flatfields and calibrates on the fly,  matches
and overplots Tycho stars (\emph{magenta}), 2MASS stars (\emph{yellow
triangles}), and displays a red cross at the requested coordinates. 
See \S\ref{sec_atv} for more details. 
\label{fig_atv}
}
\end{figure}

Using the \texttt{photoop} tools, one can calibrate the raw image
data (\texttt{idR} files) at read time, and eliminate the need to
store a second, calibrated copy of the image data (Figure
\ref{fig_atv}).  
Following is a sequence
of IDL commands to find the (run,camcol,field) for a position on the sky,
read in a calibrated image with astrometric information, match objects
in the field to the 2MASS catalog, and display the image with matches
overplotted in the IDL display tool \texttt{atv} \cite{atv}. 

\begin{verbatim}

; To see which runs cover RA=60.56, dec=0.1
IDL> imlist=sdss_findimage(60.56,0.1,rerun=137,/print)
           RA           DEC  RUN RERUN CAMCOL FIELD        XPOS        YPOS
------------- ------------- ---- ----- ------ ----- ----------- -----------
     60.56000     0.1000000  211   137      4   131      978.10      177.24
     60.56000     0.1000000  273   137      4   338      917.33      1047.2

; read the first image
file=sdss_name('idR',imlist[0].run,imlist[0].camcol,imlist[0].field, filter='r')
sdss_readimage,file,image,ivar,/reject_cr,rerun=137,hdr=hdr

; get astrometry and add to header
gsa_approx = sdss_astrom(imlist[0].run,imlist[0].camcol,imlist[0].field, filter='r')
gsssputast, hdr, gsa_approx

; display the image in the atv widget which allows panning, zooming,
;  color stretches, etc. and displays WCS coordinates from a FITS
;  header. 
atv,image,header=hdr

; Now match 2MASS stars
fobj = tmass_read(60.56, 0.1, 0.2)
astrom_adxy, gsa_approx, fobj.tmass_ra, fobj.tmass_dec, xpix=xpos, ypix=ypos

; and overplot as yellow triangles
atvplot, xpos, ypos, psym=5, symsize=1.5, color='yellow'

\end{verbatim}

The above commands demonstrate the power of some of the tools
available.  The \texttt{atv} wrapper \texttt{atvsdss} does all of the above
with the simple command

\begin{verbatim}
atvsdss, ra=60.56, dec=0.1, /catalog, rerun=137
\end{verbatim}
where the rerun number is required for use of the survey
\texttt{astrom} astrometric solutions, and \texttt{/catalog} matches
and overplots Tycho stars (\emph{magenta}), 2MASS stars (\emph{yellow
triangles}), and a red cross at the requested coordinates (Figure
\ref{fig_atv}).

\subsection{The Stellar $g-r-i$ color diagram}
\label{sec:colorcolor}

\begin{figure}[tb]
\epsscale{0.9}
\dfplot{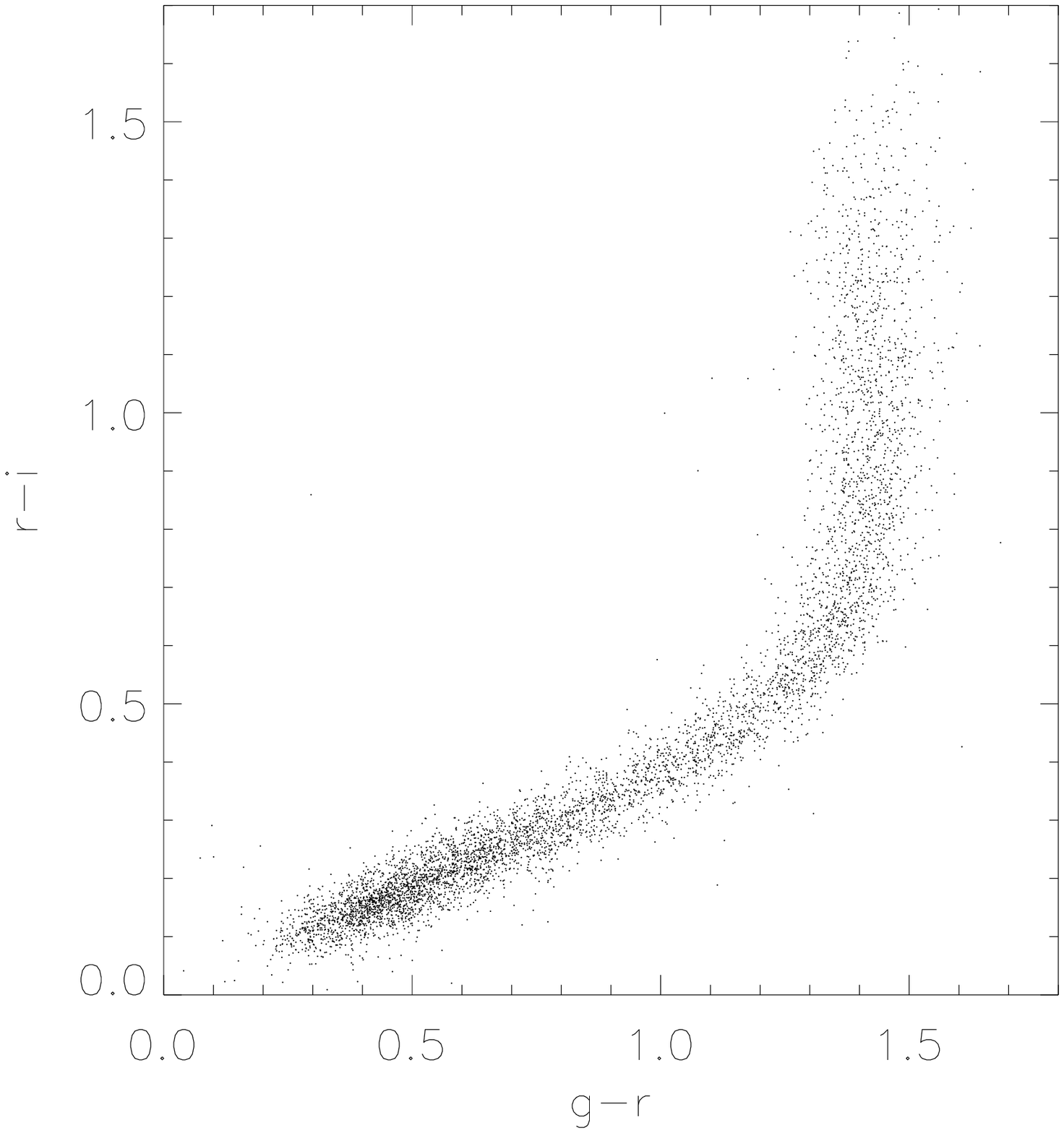}
\figcaption{A $g-r-i$ color plot showing the stellar locus, based on 
the stars in 200 fields from camera column 3 of run 273, restricted to stars 
with apparent magnitude $r < 19$. The colors are apparent colors based 
on PSF fluxes with 
no correction made for Galactic extinction. The IDL code that generated this 
figure is in 
\S \ref{sec:colorcolor}. 
\label{fig_color}
}
\end{figure}

The following is a simple example that plots the stellar locus in the 
$g-r-i$ color plane, and provides an introduction to using the catalog 
data. We start by reading in all the objects from a segment of a run, 
and then select only the stars with reliable photometry
using the object flags. We then demonstrate
the conversion of calibrated fluxes into magnitudes, and use these to
construct a magnitude-limited sample of stars. Finally, we compute and
plot the $g-r$ and $r-i$ colors of these stars. The output of this
code is shown in Figure \ref{fig_color}.

\begin{verbatim}
; Open a plot
dfpsplot, 'gricolor.ps', /square,/color

; Read in run 273, camcol 3, fields 50-250
; This reads in the data from the ``datasweep'' trimmed catalogs
objs = sweep_readobj(273, 3, rerun=137, fieldrange=[50,250])

; Select the stars -- sdss_selectobj() removes stars with these bits set:
;  (BLENDED AND NOT NODEBLEND) OR BRIGHT
; see http://www.sdss.org/dr2/products/catalogs/flags.html
; for a thorough discussion
indx = sdss_selectobj(objs, objtype='star')
stars = objs[indx]
; Eliminate all saturated stars using flags
; OBJC_FLAGS1, Bit 18 --- Saturated
; OBJC_FLAGS2, Bit 12 --- Interpolated center
; OBJC_FLAGS2, Bit 15 --- PSF Flux Interpolated

; sdss_flagval() returns 2^(flag bit set)
flag1_template = (sdss_flagval('OBJECT1','SATUR'))
flag2_template = (sdss_flagval('OBJECT2', 'INTERP_CENTER')) $
  + (sdss_flagval('OBJECT2', 'PSF_FLUX_INTERP'))

; Find all objects with none of these flags set
notset = where(((stars.OBJC_FLAGS AND flag1_template) EQ 0) $
                 AND ((stars.OBJC_FLAGS2 AND flag2_template) EQ 0))
stars = stars[notset]

; Compute the magnitudes, with a flux floor of 1.E-10 nMgy
gmag = 22.5 - 2.5*alog10(stars.psfflux[1] > 1.e-10)
rmag = 22.5 - 2.5*alog10(stars.psfflux[2] > 1.e-10)
imag = 22.5 - 2.5*alog10(stars.psfflux[3] > 1.e-10)

; Pick  stars with r < 19
w = where(rmag LT 19.0)

; Compute g-r and r-i colors
grcolor = gmag[w] - rmag[w]
ricolor = rmag[w] - imag[w]

; Plot the colors
djs_plot, grcolor, ricolor, ps=3, xr=[0,1.8], yr=[0,1.7], $
  charsize=2, xtitle='g-r', ytitle='r-i', xst=1, yst=1

; Close plot
dfpsclose
\end{verbatim}

\subsection{The Proper Motion of a Brown Dwarf}
\label{sec:browndwarf}

\begin{figure}[tb]
\epsscale{0.9}
\dfplot{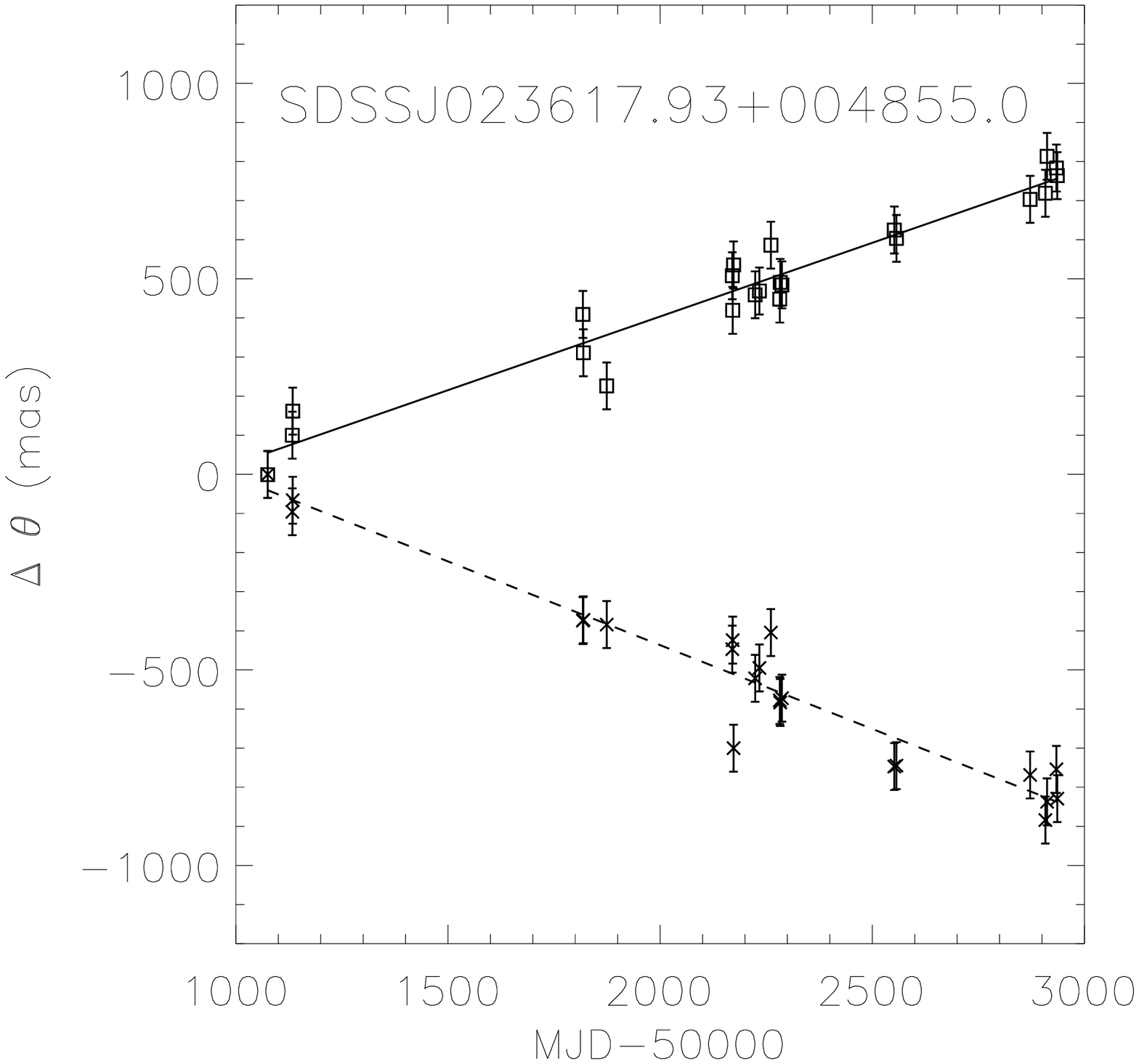}
\figcaption{The relative 
position of the L6.5 brown dwarf SDSSJ023617.93+004855.0 
as a function of time. The solid line is a fit to the change in RA (relative
to its fiducial position; \emph{squares}), while the dashed line is a
fit to the change in DEC (\emph{crosses}).
The errors in the positions are 60 mas in both coordinates. The 22
SDSS observations clearly show the proper motion
of this faint object (140 mas/year in RA, -155 mas/year in DEC). The code
that generated this figure is in
\S \ref{sec:browndwarf}.
\label{fig_bd}
}
\end{figure}

One of the features of these data are the multiple observations of
parts of the sky, enabling time domain studies. A particularly simple
example is proper motions; Figure \ref{fig_bd} shows the position of
nearby L6.5 brown dwarf \cite{schneider02,geballe02} SDSS
J023617.93+004855.0 versus time.  This object is in the SDSS Southern
Survey equatorial region and has been observed multiple times, both
during commissioning observations and during regular survey
operations.  The RA and DEC offsets with respect to its fiducial
position are shown versus time (MJD). The 22 observations span just
over five years.  The position is taken directly from the
pipeline-processed data (for a discussion of SDSS astrometry, see Pier
et al. 2003) and clearly show the proper motion of this faint object
(140 mas/year in RA, -155 mas/year in DEC).
The position errors determined from this fit are 60 mas in each
coordinate.
Several SDSS brown dwarfs were identified in the commissioning and
calibration data (Fan \etal\ 2000) and a project to measure the proper
motions of nearby stars and substellar objects is underway.

The following code fragment creates Figure \ref{fig_bd}; in particular,
it demonstrates the use of \texttt{sdss\_findallobj} to find multiple
observations of objects in the data.  Most observations of this object
are not yet public; we include them in the figure to show what will 
be possible with the complete data set. 

\begin{verbatim}
;Open the plot
dfpsplot, 'bd.ps', /color,/square

; This is SDSSJ023617.93+004855.0
ra = 39.0747106
dec = 0.8152421

; Find all the observations of this object in rerun 137
imlist = sdss_findallobj(ra, dec, rerun=137)

; Specify the fields to extract
select_tags = ['RUN', 'CAMCOL', 'FIELD', 'ID', 'MJD', $
               'PSFFLUX*', $
               'RA', 'DEC', 'OFFSET*', 'CALIB_STATUS']

; Read in these fields
objs = sdss_readobjlist(inlist=imlist, select_tags=select_tags, /silent)

; Define time axis
dmjd = objs.mjd - 50000

; Compute the change in RA and DEC in mas
dra = (objs.ra - ra)*3.6d6*cos(dec/!radeg)
ddec = (objs.dec - dec)*3.6d6


; Plot the change in RA with time
djs_plot, dmjd, dra, yr=[-1200,1200], yst=1, charsize=2, $
  thick=3, ps=6, xtitle='MJD-50000', ytitle='\Delta \theta (mas)'
; Plot typical error of 60 mas
errplot, dmjd, dra-60.0, dra+60.0, thick=3

; Plot the change in DEC with time
djs_oplot, dmjd, ddec, thick=3, ps=7, line=2
; Plot error of 60 mas
errplot, dmjd, ddec-60.0, ddec+60.0, thick=3

; Overplot linear fits
oplot,dmjd,poly(dmjd,linfit(dmjd, dra)), thick=3
oplot,dmjd,poly(dmjd,linfit(dmjd, ddec)), thick=3, line=2

; Output a label
djs_xyouts, 1100, 900, 'SDSSJ023617.93+004855.0', charsize=2.5

; Close the plot
dfpsclose
\end{verbatim}                     

\subsection{Star Formation and Young Stellar Objects}
\label{sec_yso}

Figure \ref{fig_orion}
shows a color composite from the three most sensitive SDSS bands
(g, r and i) of a region from runs 259/273 centered on the star forming
region NGC 2068/NGC 2071/HH24-26.  Note the line of Herbig-Haro (HH)
objects in the outflow jet.  As their emission is dominated by
emission lines, HH objects have 
characteristic colors and can be automatically identified and counted
in these regions of high obscuration.  We find that there is a very 
different density distribution of these objects in the Taurus and Orion
star-forming regions (Knapp et al., 2004, in preparation).

The spectral energy distributions of the 
stars in NGC 2068 and NGC 2071 can be constructed
from the five SDSS and three 2MASS bands, giving the luminosity 
function, the frequency of low mass objects and the incidence of
circumstellar dust. UV-excess objects, i.e. young T Tauri stars, 
can be discovered and mapped across the Orion and Taurus 
star-forming regions (McGehee \etal\ 2004).

\subsection{Structure of the Galactic Halo and Vertical Structure of the Disk}

Because of the depth and color discrimination of SDSS photometry, the
SDSS data have enabled major contributions to studies of the structure
of the Galactic halo \cite{yanny00, ivezic00, oden01, newberg03a}.
These and other studies of distant halo objects 
\cite{ibata95,johnston95,majewski03,martinez04}
show the
presence of long streams in the halo produced by the tidal disruption
of satellite systems (small galaxies and loosely bound globular
clusters). The data described herein allow these streams to be traced
much closer to the Galactic plane. 
The SDSS stellar data at low Galactic latitudes can be 
used to investigate the relationships among the thin disk, the thick
disk and the Galactic halo, tracing the metallicity of each component.
The discovery of these halo structures, and
the ability of SDSS to follow the structure close to the Galactic
plane, provide the impetus for a new initiative to use the SDSS
hardware and software to map the structure of the Galactic halo and the
disk-halo interface, known as SEGUE \cite[Sloan Extension for Galactic
Underpinnings and Evolution]{newberg03b}.

The vast majority of the red stars detected by SDSS are late-type disk 
dwarfs and subdwarfs. Hawley et al. (2002, and \emph{in prep.})
show a well-established 
color-absolute magnitude relationship for dwarfs of spectral type M0
and later, which allows the vertical mapping of the disk to a distance of about
1 kpc from the Galactic plane, as well as the three-dimensional distribution of
the interstellar dust.  Measurements at different Galactic latitudes,
especially at low latitudes, allow the measurement of the disk scale
length and disk scale height for these stars \cite{mcgehee04b, juric04}.

\subsection{Properties of Interstellar Dust}

Reddening measured by fitting the position of the ``blue tip'' of the
stellar locus in ($u,g,r,i,z$) color space can be used to investigate
possible deviations from the standard extinction law by comparing $E(u-g)$
with $E(g-r)$ in different directions (Finkbeiner et al. 2004).  The
mosaic images in Figures \ref{fig_orion}, \ref{fig_taurus} and
\ref{fig_cygnus} demonstrate that the dust clouds
also have different emission/reflection properties. In particular, the
Taurus dark clouds (Fig. \ref{fig_taurus}) glow at red wavelengths, perhaps
due to extended red emission \cite{witt98}.  Differences between
the emission properties of dust clouds, and of the illuminating starlight,
can be investigated over large areas with these data.


\section{Acknowledgments}

Partial support for the computer systems required to process, store and 
distribute these data was provided by NASA via grant NAG5-6734, by
NASA's Hubble Fellow program,
and by Princeton University. The
NASA funds were originally provided in support of an Associate Investigator
Program with WIRE.  We thank NASA and the WIRE PI, Perry Hacking, both
for the opportunity to work in the WIRE team and for generously permitting
us to apply the funding towards support of the Orion data reduction.
DPF is a Hubble Fellow supported by HST-HF-00129.01-A.
C. Rockosi is a Hubble Fellow supported by HST-HF-01143.01-A.
We also thank Princeton University for generous support.
This research made use of the IDL Astronomy User's Library at
Goddard Space Flight
Center\footnote{http://idlastro.gsfc.nasa.gov/}.

This publication makes use of data products from the Two Micron All
Sky Survey, which is a joint project of the University of
Massachusetts and the Infrared Processing and Analysis
Center/California Institute of Technology, funded by the National
Aeronautics and Space Administration and the National Science
Foundation.

Funding for the SDSS has been
provided by the Alfred P. Sloan Foundation, the Participating
Institutions, the National Aeronautics and Space Administration, the
National Science Foundation, the U.S.  Department of Energy, the
Japanese Monbukagakusho, and the Max Planck Society. The SDSS Web site
is http://www.sdss.org/.

The SDSS is managed by the Astrophysical Research Consortium (ARC) for
the Participating Institutions. The Participating Institutions are The
University of Chicago, Fermilab, the Institute for Advanced Study, the
Japan Participation Group, The Johns Hopkins University, Los Alamos
National Laboratory, the Max-Planck-Institute for Astronomy (MPIA),
the Max-Planck-Institute for Astrophysics (MPA), New Mexico State
University, University of Pittsburgh, Princeton University, the United
States Naval Observatory, and the University of Washington.

\newpage

\clearpage
\appendix
\section{Structure of the \texttt{calibObj} Files}
\label{app_calibobj}

A current description of the \texttt{calibObj} data format is
available at the website$^1$ and is reproduced here for convenience. 
Fields unchanged from the \texttt{fpObjc} files are not repeated here,
and are described in Stoughton et al. (2002).

\scriptsize
\begin{verbatim}
1. calibObj book-keeping values:

   RUN             LONG      Run number
   RERUN           STRING    Rerun name
   CAMCOL          LONG      Camera column [1..6]
   FIELD           LONG      Field number
   ID              LONG      Object ID, starting at 1 and unique in each field
   PARENT          LONG      ID of parent, -1 if this is a parent object
   NCHILD          LONG      Number of children deblended from this object
   MJD             LONG      Modified Julian Day for date of observation
   TAI             DOUBLE[5] Mean time of observation for the object center
                             in each filter, where TAI = 24 * 3600 * MJD
   AIRMASS         FLOAT[5]  Airmass approximated as csc(zenith-angle)
   PSP_STATUS      LONG[5]   "Status" field from the psField file
   CALIB_STATUS    INT[5]    Status of photometric calibration

2. calibObj calibration parameters:

   NMGYPERCOUNT    FLOAT[5]    Calibration of nano-maggies per count
   NMGYPERCOUNT_IVAR  FLOAT[5] Formal error in the photometric calibration as
                               an inverse variance; =0 if there were no calibration
                               stars and a default calibration was used instead
   CLOUDCAM        INT[5]      Photometricity in each filter:
                                -1=unknown, 0=cloudy, 1=clear
   EXTINCTION      FLOAT[5]    Galactic extinction in magnitudes, defined as
                               [5.155, 3.793, 2.751, 2.086, 1.479] * E(B-V)
                               where E(B-V) is from the SFD98 dust maps.
                               Fluxes (and their errors) can be corrected
                               for Galactic extinction as follows:
                                 Corrected-Flux = Flux * 10^(Extinction/2.5)
   PIXSCALE        FLOAT[5]    Pixel scale [arcsec/pix]
   PSF_FWHM        FLOAT[5]    PSF FWHM [arcsec]
   PHI_OFFSET      FLOAT[5]    Calibration angle [degrees] to add to any
                               uncalibrated position angles such that they mean
                               degrees east of north.

3. calibObj structure calibrated quantities:

   RA              DOUBLE      J2000 RA [deg] in canonical (r-band) filter
   DEC             DOUBLE      Declination [deg] in canonical (r-band) filter
   OFFSETRA        FLOAT[5]    RA of center in this filter relative to RA [arcsec]
   OFFSETDEC       FLOAT[5]    DEC of center in this filter relative to RA [arcs
   PHI_ISO_DEG     FLOAT[5]    Isophotal fit, angle E of N [deg]
   PHI_DEV_DEG     FLOAT[5]    De Vaucouleurs fit, angle E of N [deg]
   PHI_EXP_DEG     FLOAT[5]    Exponential fit, angle E of N [deg]
   SKYFLUX         FLOAT[5]    Sky level [nMgy/arcsec^2]
   SKYFLUX_IVAR    FLOAT[5]
   PSFFLUX         FLOAT[5]    PSF flux [nMgy]
   PSFFLUX_IVAR    FLOAT[5]
   FIBERFLUX       FLOAT[5]    Fiber flux, 3 arcsec diameter [nMgy]
   FIBERFLUX_IVAR  FLOAT[5]
   MODELFLUX       FLOAT[5]    Model flux [nMgy]
   MODELFLUX_IVAR  FLOAT[5]
   PETROFLUX       FLOAT[5]    Petrosian flux [nMgy]
   PETROFLUX_IVAR  FLOAT[5]
   DEVFLUX         FLOAT[5]    De Vaucouleurs flux [nMgy]
   DEVFLUX_IVAR    FLOAT[5]
   EXPFLUX         FLOAT[5]    Exponential flux [nMgy]
   EXPFLUX_IVAR    FLOAT[5]
   APERFLUX        FLOAT[NAPER,5]  Aperture fluxes [nMgy]in radii of size 0.223,
                                   0.670,1.024,1.745,2.972,4.584,7.359,11.306,
                                   18.020,27.922,43.770,68.307,106.73,166.52,
                                   260.37 arcsec; our default is to return
                                   only those fluxes out to 11.3 arcsec,
                                   unless NAPER is explicitly specified
   APERFLUX_IVAR   FLOAT[NAPER,5]

\end{verbatim}
\normalsize

\section{Using Inverse Variance}
\label{app_ivar}

Flux errors are expressed as inverse variance (ivar) in the
\texttt{calibObj} files.  This is convenient for inverse variance
weighting of combined quantities, and handles zero signal/noise
(ivar = 0) gracefully, as the following example demonstrates. 
\subsection{Color Cut}
To select all stars with flux ratio $r/i$ greater than some color
ratio $c$ with $n\sigma$ confidence, one writes the inequality as
$r-ic > n\sigma$, where $\sigma$ is the uncertainty
in the quantity $r-ic$, or
$$
\sigma=\left(\frac{1}{I_r}+\frac{c^2}{I_i}\right)^{1/2}
$$
where $I_r$ and $I_i$ are the r and i-band inverse variance respectively. 
Multiplying by $\sqrt{I_rI_i}$ yields the inequality
$$
(r-ic)\sqrt{I_rI_i} > n\sqrt{I_i+c^2I_r}
$$
which concisely expresses the desired condition without a division by
any of the parameters, and correctly handles the limiting cases of
zero or negative fluxes, or inverse variance equal to zero
(non-measurement).  The conventional expression of this with
magnitudes and $\sigma$ instead of inverse variance requires handling
the special cases of zero or negative fluxes in either band, and
cannot tolerate zero flux in either band.

Many astronomers will want to work with magnitudes, related
to nMgy fluxes by 
$$
m = 22.5 - 2.5\log_{10}(flux)
$$
but will find that any flux and color cuts are most easily performed
in flux units.  These fluxes and magnitudes are on the natural SDSS
system, and are offset from the true AB system by a few hundredths of
a magnitude in each band (see Abazajian et al. 2004 for estimated offsets).

\subsection{Extinction Corrections}
Extinction values are presented in magnitudes for historical reasons. 
The extinction estimates provided are based on maps of IR dust
emission (Schlegel et al. 1998) and are most reliable in the limit of long
wavelengths.  Extinction values are given in the \texttt{calibObj}
files for each SDSS band, assuming
a standard interstellar reddening law of $R_V = 3.1$ in the
parameterization of Cardelli, Clayton, \& Mathis (1989).  
Because the SFD98 dust map is calibrated to $E(B-V)$, it is presented
as $E(B-V)$ reddening.  This has led to the misunderstanding that the
presented $E(B-V)$ is valid for all values of $R_V$, when in fact it
is the the near IR extinction (e.g. SDSS z band) that is roughly
constant with changes in $R_V$.  Therefore, if the dust along a given
line of sight is known to have a different value of $R_V$, it is
necessary to extrapolate the extinction from a red band (e.g. z band)
to the desired band using the appropriate reddening law.

The reader is reminded that flux uncertainties must be modified when
correcting for extinction.  Unlike magnitude uncertainties, which are
fractional uncertainties, flux uncertainties are absolute, and must be 
amplified by the same factor as the flux when extinction correction
are applied.  To correct flux $F$ and inverse variance $I$ for an
extinction of $A$ magnitudes, use:
$$
F_{corr}=10^{0.4 A} F
$$ 
and $$
I_{corr}=10^{-0.8 A} I.
$$


\bibliographystyle{unsrt}
\bibliography{gsrp}

\begin{thebibliography}{DUM}

\bibitem[Abazajian et al. 2003]{abazajian03}
Abazajian, K., Adelman-McCarthy, J.~K., Ag{\" u}eros, M.~A., et
al. 2003, \aj, 126, 2081 (DR1)

\bibitem[Abazajian et al. 2004]{abazajian04}
Abazajian, K., Adelman-McCarthy, J.~K., Ag{\" u}eros,
M.~A., et al. 2004, AJ in press (DR2)

\bibitem[Barth 2001]{atv}
Barth, A. J. 2001, in ASP Conf. Ser., Vol. 238,
Astronomical Data Analysis Software and Systems X,
eds.  F. R. Harnden, Jr., F. A. Primini, \& H. E. Payne
(San Francisco: ASP), 385

\bibitem[Becker \etal\ 1995]{becker95}
Becker, R.H., White, R.L., Helfand, D.J. 1995, \apj, 450, 559

\bibitem[Cardelli{,} Clayton \& Mathis 1989]{ccm89}
Cardelli, J. A., Clayton, G. C., \& Mathis, J. S. 1989, \apj, 345, 245

\bibitem[Cutri \etal\ 2003]{cutri03}
Cutri, R. M., et al. 2003, Explanatory Supplement to the 2MASS All Sky
Data Release (Pasadena: Caltech) and VizieR Online Data Catalog, 2246

\bibitem[Fan \etal\ 2000]{fan00}
Fan, X., Knapp, G., Strauss, M. A., \etal\ 2000, AJ, 119, 928

\bibitem[Finkbeiner \etal\ 2004]{bluetip}
Finkbeiner, D. P. \etal\ 2004, in prep. 

\bibitem[Fukugita et al. 1996]{fukugita96}
Fukugita, M., Ichikawa, T., Gunn, J.E., Doi, M., Shimasaku, K., \& 
 Schneider, D.P. 1996, \aj, 111, 1748

\bibitem[Geballe \etal 2002]{geballe02}
Geballe, T.R., Knapp, G.R., Leggett, S.K., et al. 2002, ApJ, 564, 466

\bibitem[Gunn et al. 1998]{gunn98}
Gunn, J.E., Carr, M., Rockosi, C.M., Sekiguchi, M., et al. 1998, AJ, 116, 
 3040

\bibitem[Hawley \etal 2002]{hawley02}
Hawley, S.L., Covey, K.R., Knapp, G.R., et al. 2002, AJ, 123, 3409

\bibitem[H{\o}g \etal 2000]{hog00}
H{\o}g, E.,  Fabricius, C., Makarov, V.V., et al. 2000, \aap, 355, L27 

\bibitem[Hogg et al. 2001]{hogg01}
Hogg, D. W., Finkbeiner, D. P., Schlegel, D. J. \& Gunn, J. E. 2001, 
AJ, 122, 2129

\bibitem[Ibata \etal 1995]{ibata95}
Ibata, R.A., Gilmore, G., \& Irwin, M.J. 1995, MNRAS, 277, 781

\bibitem[Ivezi\'c \etal 2000]{ivezic00}
Ivezi\'c, \v{Z.}, Goldston, J., Finlator, K., et al. 2000, AJ, 120, 963

\bibitem[Johnston \etal 1995]{johnston95}
Johnston, K.V., Spergel, D.N., \& Hernquist, L. 1995, ApJ, 451, 598

\bibitem[Juri\'c \etal 2004]{juric04}
Juri\'c, M., Ivezi\'c, \v{Z.}, et al. 2004, in prep.

\bibitem[Kniazev \etal\ 2003]{kniazev03}
Kniazev, A. Y., Grebel, E. K., Hao, L., Strauss, M. A.,
Brinkmann, J., \& Fukugita, M. 2003, ApJ, 593, L73

\bibitem[Lupton \etal\ 1999]{lupgunn}
Lupton, R.H., Gunn, J.E., \& Szalay, A.S. 1999, AJ, 118, 1406

\bibitem[Lupton \etal\ 2001]{lupton01}
Lupton, R.H., Gunn, J.E., Ivezi\'c, \v{Z}., Knapp, G.R., Kent, S.M., \& 
 Yasuda, N. 2001, ADASS X, ed. F.R. Harnden, F.A. Primini, \& H.E.
 Payne, ASP Conf. Proc. 238, 269

\bibitem[Lupton \etal\ 2003]{lupton03}
Lupton, R.H., Ivezi\'c, \v{Z}., Gunn, J.E., Knapp, G.R., Strauss, M.A., \& 
 Yasuda, N. 2003, Proc. SPIE, 4836, 350

\bibitem[Majewski \etal 2003]{majewski03}
Majewski, S., Skrutskie, M.F., Weinberg, M.D., \& Ostheimer, J.C. 
 2003, ApJ, 599, 1082

\bibitem[Mart\'{\i}nez-Delgado \etal 2004]{martinez04}
Mart\'{\i}nez-Delgado, D., G\'omez-Flechoso., M., Aparacio, A., \&
 Carrera, R. 2004, ApJ, 601, 242

\bibitem[McGehee \etal\ in prep]{mcgehee04}
McGehee, P. \etal\ 2004, in preparation

\bibitem[McGehee \etal\ in prep]{mcgehee04b}
McGehee, P.M., Anderson, K.S.J., Hobbs, L.M., \& York, D.G., in preparation

\bibitem[Monet \etal\ 2003]{monet03}
Monet, D.G, Levine, S.E., Canzian, B., et al. 2003, \aj, 125, 984

\bibitem[Newberg \etal 2003a]{newberg03a}
Newberg, H.J., Yanny, B., Grebel, E., et al. 2003a, ApJ, 596, L191

\bibitem[Newberg \etal 2003b]{newberg03b}
Newberg, H.J., \& SDSS Collaboration 2003b, AAS, 203, 112.11

\bibitem[Odenkirchen \etal 2001]{oden01}
Odenkirchen, M., Grebel, E., Rockosi, C., et al. 2001, ApJ, 548, L165

\bibitem[Oke \& Gunn 1983]{oke83}
Oke, J. B. \& Gunn, J. E. 1983, \apj, 266, 713

\bibitem[Pier et al. 2003]{pier03}
Pier, J.R., Munn, J.A., Hindsley, R.B., Hennessy, G.S., Kent, S.M., Lupton,
 R.H., \& Ivezi\'c, \v{Z.} 2003, \aj, 125, 1559

\bibitem[Schlegel et al. 2004]{schlegel04} 
Schlegel, D. J. \etal\ 2004, in prep.

\bibitem[Schlegel{,} Finkbeiner \& Davis 1998]{sfd98} 
Schlegel, D. J., Finkbeiner, D. P., \& Davis M. 1998, \apj, 500, 525 (SFD98)

\bibitem[Schneider \etal 2002]{schneider02}
Schneider, D.P., Knapp, G.R., Hawley, S.L., et al. 2002, AJ, 123, 458

\bibitem[Smith et al. 2002]{smith02}
Smith, J.A., Tucker, D.L., Kent, S.M., et al. 2002, \aj, 123, 2121

\bibitem[Stoughton et al. 2002]{stoughton02}
Stoughton, C., Lupton, R.H., Bernardi, M., et al. 2002, \aj, 123, 485 (EDR)

\bibitem[Witt \etal 1998]{witt98}
Witt, A.N., Gordon, K.D., \& Furton, D.G. 1998, ApJ, 501, L111

\bibitem[Yanny \etal 2000]{yanny00}
Yanny, B., Newberg, H.J., Kent, S., et al. 2000, ApJ, 540, 825

\bibitem[York et al. 2000]{york00}
York, D.G., Adelman, J., Anderson, J., et al. 2000, \aj, 120, 1579


\bibitem[Zacharias \etal 2000]{zach00}
Zacharias, N., Rafferty, T.J., Zacharias, M.I. 2000, ADASS IX, 
ed. N. Manset, C. Veillet, and D. Crabtree, ASP Conf. Ser. 216, 427

\end{thebibliography}

\clearpage

\begin{figure}[tb]
\figcaption{Orion: a section of runs 259 and 273 centered at J2000
$(\alpha,\delta) = (05^h48^m28^s, +00\degree04'48'').$
The
(R,G,B) color planes represent SDSS (i,r,g) filters
respectively, so that \Halpha\ emission appears green.
Celestial North is indicated by the arrow in (d). 
The widths of panels ($a-d$) are $108', 54', 27' and 13.5'$ respectively. 
Nebulae NGC 2071 (\emph{upper}) and NGC 2068 are visible in panel
($b$).  Panel ($d$) zooms in on the object Herbig-Haro 24 (\emph{upper
 middle}).
\label{fig_orion}
}
\end{figure}

\begin{figure}[tb]
\figcaption{Taurus: a section of runs 3512 and 3559 
centered at $(\alpha,\delta) = (04^h24^m40^s, +25\degree41'24'')$.
The color scheme and scale are identical to Figure \ref{fig_orion}. 
The large dark cloud in the lower right of ($a$) is Barnard 215, which
joins with Lynds 1506 further to the right.  Panel ($d$) zooms in on
Herbig-Haro 31A,B,C,D, and associated outflows (\emph{middle left}).
\label{fig_taurus}
}
\end{figure}

\begin{figure}[tb]
\figcaption{Cygnus: a section of runs 4115 and 4119 centered at
$(\alpha,\delta) = (20^h36^m04^s, +40\degree26'24'')$. 
The color scheme and scale are identical to Figure \ref{fig_orion}. 
The \HII\ region LBN 271 is visible in panel ($a$)
(\emph{middle left}).  Panel ($d$) zooms in on a beautiful complex of
dark clouds in front of LBN 258.
\label{fig_cygnus}
}
\end{figure}

\clearpage
\begin{deluxetable}{rrrcrrcrrrrrc}
\tabletypesize{\scriptsize}
\tablewidth{0pt}
\tablecaption{SDSS Galactic Plane Runs
   \label{table_conversions}
}
\tablehead{
\colhead{Run} &
\colhead{Date} &
\colhead{MJD} &
\colhead{Strip} &
\colhead{Node} &
\colhead{Incl} &
\colhead{$F_0$} &
\colhead{$F_1$} &
\colhead{$A$} &
\colhead{$l$} &
\colhead{$b$} &
\colhead{PSF} &
\colhead{Phot}
\\
\colhead{} &
\colhead{UT} &
\colhead{} &
\colhead{} &
\colhead{deg} &
\colhead{deg} &
\colhead{} &
\colhead{} &
\colhead{deg$^2$} &
\colhead{deg} &
\colhead{deg} &
\colhead{$''$} &
\colhead{}
\\
\colhead{(1)} &
\colhead{(2)} &
\colhead{(3)} &
\colhead{(4)} &
\colhead{(5)} &
\colhead{(6)} &
\colhead{(7)} &
\colhead{(8)} &
\colhead{(9)} &
\colhead{(10)} &
\colhead{(11)} &
\colhead{(12)} &
\colhead{(13)}
}
\startdata
   211&  1998/10/29&  51115& 82S& 283.22&   0.01&   11&  385&   71.2&  210.2&   -3.3& 1.36& Y \\
   250&  1998/11/16&  51133& 82N&  62.10&   0.02&   11&  455&   84.5&  201.9&  -18.0& 1.55& Y \\
   259&  1998/11/17&  51134& 82N& 299.41&   0.01&   12&  567&  105.6&  206.7&   -9.5& 1.19& Y \\
   273&  1998/11/19&  51136& 82S& 286.54&   0.01&   11&  534&   99.6&  206.1&  -11.1& 1.41& N \\
   297&  1998/11/22&  51139& 82O&  92.04&   0.04&   11&  203&   36.7&  206.0&  -11.0& 1.47& Y \\
   307&  1998/11/23&  51140& 82N& 318.93&   0.01&   11&  186&   33.4&  212.4&    0.0& 1.21& N \\
   308&  1998/11/23&  51140& 10N& 271.30&   0.01&   11&  216&   39.1&  216.0&    4.2& 1.24& Y \\
   994&  1999/10/06&  51457& 76S& 275.04&  15.00&   11&  133&   23.4&  209.9&   -5.0& 1.84& N \\
  1923&  2000/12/08&  51886&  0O& 215.00&  44.99&   29&   71&    8.2&   86.4&    0.0& 1.75& U \\
  1924&  2000/12/08&  51886&  0O& 354.00&  45.99&   34&   78&    8.6&  168.6&   11.9& 1.88& U \\
  1925&  2000/12/08&  51886&  0O& 349.98&  45.78&   17&   48&    6.1&  165.3&    6.6& 1.60& U \\
  2955&  2002/02/07&  52312& 82N&   9.82&   0.00&   21&  166&   27.7&  209.3&   -4.5& 2.07& Y \\
  2960&  2002/02/08&  52313& 82S&  25.79&   0.01&   11&  127&   22.2&  208.0&   -7.4& 1.16& Y \\
  2968&  2002/02/09&  52314& 82N&  90.64&   0.00&   19&   94&   14.4&  207.8&   -7.3& 1.68& Y \\
  3511&  2002/12/06&  52614&  0O&  65.19&  90.14&   14&   89&   14.4&  163.5&   -8.2& 1.17& Y \\
  3512&  2002/12/06&  52614&  0O& 335.41&  25.32&   11&   90&   15.2&  177.3&   -9.3& 1.22& Y \\
  3557&  2002/12/31&  52639&  0O&  65.50&  89.96&   11&   87&   14.6&  165.6&  -10.5& 1.26& Y \\
  3559&  2002/12/31&  52639&  0O& 335.41&  25.31&   11&   86&   14.4&  176.4&   -9.9& 1.20& Y \\
  3610&  2003/01/27&  52666& 61N& 274.99&  52.50&   17&  148&   25.1&  139.0&  -10.8& 1.63& Y \\
  3628&  2003/01/28&  52667& 61S& 275.01&  52.52&   17&  165&   28.3&  137.3&  -10.9& 0.97& Y \\
  3629&  2003/01/28&  52667& 61N& 275.00&  52.50&   11&  129&   22.6&  142.0&  -10.9& 0.97& Y \\
  3634&  2003/01/29&  52668& 62N& 275.00&  50.01&   18&  162&   27.5&  159.1&  -13.0& 1.04& Y \\
  3642&  2003/02/01&  52671& 62S& 274.99&  49.99&   11&   97&   16.5&  135.6&  -13.2& 1.73& Y \\
  3643&  2003/02/01&  52671& 62S& 274.99&  50.00&   11&  119&   20.7&  159.1&  -13.2& 1.68& Y \\
  4114&  2003/09/20&  52902&  0S& 309.49&  90.02&   41&  127&   16.5&   80.1&    0.0& 1.59& Y \\
  4115&  2003/09/20&  52902&  0S& 309.34&  90.15&   11&   69&   11.2&   78.7&    0.0& 1.36& Y \\
  4116&  2003/09/20&  52902&  0O& 219.49&  41.40&   11&   77&   12.7&   82.1&    0.0& 1.40& Y \\
  4119&  2003/09/20&  52902&  0O& 309.40&  90.08&   11&   70&   11.4&   79.8&    0.0& 1.71& Y \\
\enddata
\tablecomments{
Col. (1): SDSS Run number. 
Col. (2): UT Date.
Col. (3): UT MJD.
Col. (4): Survey strip number, 82 and 10 are on the Celestial Equator; 
             N=north, S=south, O=other (non-survey strips).
Col. (5,6): J2000 node, inclination of stripe great circle. 
Col. (7): Start field for run -- in some cases ramp up takes more than
11 fields.
Col. (8): End field.
Col. (9): Gross area in square degrees. 
Col. (10, 11): Galactic $(l,b)$ of closest approach to the Galactic
Plane of any field center.
Col. (12): Median PSF FWHM (r-band, camera column 3).
Col. (13): Photometric?  (Y=yes, N=no, U=unkown). 
Run 273 has an aperture obstruction early in the run, and run 307 is
slightly cloudy near the end.  Data in the 1998-1999 runs are now public. 
\label{tab_runlist}
}
\end{deluxetable}
\begin{deluxetable}{lccccl}
\tabletypesize{\scriptsize}
\tablecaption{CVS Products}
\tablewidth{0pt}
\tablehead{
\colhead{Product} &
\colhead{CVS Version} &
\colhead{Language} &
\colhead{Analysis?} &
\colhead{Description}
}
\startdata
\texttt{dust} &  v0\_0 & IDL/Fortran/C & Opt & The Schlegel et al. (1998) \\
& & & & $E(B-V)$ extinction maps \\  
\texttt{eups} & v0\_4 & Perl & Opt & Product management software \\
\texttt{first} & v03Apr11 & N/A & Opt & The FIRST \cite{becker95} Radio catalog\\
\texttt{idlutils} & v5\_0\_0 & IDL/Fortran/C & Yes & General IDL tools; {\it Required for all IDL tasks} \\
\texttt{photoop} & v1\_1 & Perl/IDL & Yes & SDSS Perl/IDL tools \\
\texttt{runList.par} & N/A & N/A & Yes & The list of runs processed, and their locations \\
\texttt{twomass} & allsky & N/A & Opt & The 2-MASS \cite{cutri03} \\
& & & & Point Source Catalog \\
\texttt{tycho2} & v0\_0 & N/A & Opt & The Tycho \cite{hog00} astrometric catalog \\
\texttt{ucac} & v2\_final & N/A & Opt & The UCAC \cite{zach00} astrometric \\
& & & & catalog \\
\texttt{usno} & vB & N/A & Opt & The USNO-B \cite{monet03} astrometric \\
& & & &catalog \\
\enddata
\tablecomments{
The list of all products used for either processing, analysing or
accessing the data. We limit ourselves to those products unique to
this reduction; the reader is referred to Stoughton et al. (2002) for
the official survey pipeline code. The column ``Analysis'' marks products
either necessary (``Yes'') or optional (``Opt'') for data analysis.
\label{tab:products}
}
\end{deluxetable}
\clearpage

\end{document}